\documentclass[10pt,onecolumn]{article}
\usepackage{joss}

\usepackage[style=apa, backend=biber]{biblatex}
\jossRepoURL{https://github.com/DataScribe-Cloud/datascribe\_api}
\jossArchiveDOI{10.5281/zenodo.17090844}
\jossDocsURL{https://datascribe-cloud.github.io/datascribe_api/}
\jossDate{Jun 2026}
\jossLogo{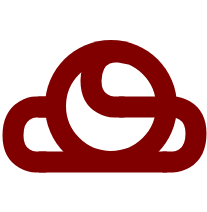}
\jossFooter{Sarıtürk, Arróyave \& Attari (2026). \textit{datascribe\_api}: Enabling Data-Driven Materials Discovery with DataScribe}

\jossauthor[1]{Doğuhan Sarıtürk}
\jossorcid{0000-0003-0008-1165}
\corrauthor{sariturk@tamu.edu}

\jossauthor[1,2,3]{Raymundo Arróyave}
\jossorcid{0000-0001-7548-8686}

\jossauthor[1]{Vahid Attari}
\jossorcid{0000-0003-3507-8847}

\affiliation[1]{Department of Materials Science and Engineering, Texas A\&M University, USA}
\affiliation[2]{J. Mike Walker '66 Department of Mechanical Engineering, Texas A\&M University, USA}
\affiliation[3]{Wm Michael Barnes '64 Department of Industrial and Systems Engineering, Texas A\&M University, USA}

\title{datascribe\_api: Enabling Data-Driven Materials Discovery with DataScribe}

\begin{document}

\sloppy

\section{Summary}

datascribe\_api is a Python library and command-line interface (CLI) that gives researchers programmatic access to DataScribe Cloud \parencite{singh2026}, an AI-native platform for materials design and discovery. DataScribe integrates two data services: user-managed scientific data tables, where research groups store experimental measurements, simulation outputs, and derived features; and curated materials science databases from the Materials Project \parencite{Horton2025,jain2013}, AFLOW \parencite{curtarolo2012,rose2017}, and OQMD \parencite{saal2013,kirklin2015}. datascribe\_api exposes both through a single, typed Python interface, so researchers can retrieve, filter, and analyze data without writing custom HTTP code.

The library handles authentication, connection management, and transient network errors automatically. It exposes nine query endpoints as ordinary Python method calls. Filter conditions use standard Python comparison syntax and are passed as a list, keeping query logic readable without format-specific boilerplate. Pydantic \parencite{pydantic2024} models validate all API responses at parse time, so schema mismatches surface as exceptions rather than silent wrong values. Each result collection converts to a pandas \parencite{mckinney2010,reback2020} DataFrame in a single call, so retrieved data is directly available for machine learning and statistical analysis. Every materials search result carries a provenance field identifying its source database, satisfying FAIR traceability requirements.

The library includes a CLI that exposes all nine endpoints as subcommands, using Typer \parencite{ramirez2019} for argument parsing and Rich \parencite{willmott2020} for terminal output. Every subcommand supports machine-readable JSON output for shell pipelines and LLM agents. Pre-built wheels for Python~3.11 and later are published on PyPI under the GPLv3 license for Linux, macOS, and Windows.

\section{Statement of Need}

datascribe\_api is intended for computational materials scientists, experimentalists who analyze data programmatically, and research software engineers who build reproducible data pipelines. It works equally well in interactive Jupyter notebooks and in automated workflows feeding data to machine learning models or Bayesian optimizers.

The Materials Project \parencite{jain2013}, AFLOW \parencite{curtarolo2012,rose2017}, and OQMD \parencite{saal2013,kirklin2015} are among the most widely used open materials databases, yet each has its own Python client with a distinct authentication scheme, query syntax, and response structure. Researchers comparing results across databases must maintain three separate packages with incompatible interfaces and write reconciliation code to bring results into a common format. That overhead discourages thorough cross-database validation and makes pipelines harder to share.

Beyond public databases, most research groups keep experimental measurements or simulation outputs in spreadsheets or local databases with no programmatic interface. Getting that data into an analysis script means writing bespoke authentication and parsing code from scratch, once per project.

DataScribe provides AI-driven capabilities including uncertainty-aware Bayesian optimization, surrogate model training, and agentic data workflows for closed-loop materials discovery. These features consume data from both the materials databases and the user-managed tables. datascribe\_api exposes those sources to Python-based workflows, so the full DataScribe stack can be embedded in scripted, reproducible pipelines without requiring the web UI.

\section{State of the Field}

\texttt{pymatgen} \parencite{ong2013} is the standard library for querying the Materials Project. It provides rich analysis functionality alongside data retrieval, which makes it valuable for MP-centric workflows but less suited as a generic query client: its architecture is built around materials analysis, not retrieval, and it does not cover AFLOW, OQMD, or user-managed data tables. AFLOW provides access to its repository through the AFLUX REST API \parencite{taylor2014}, a URL-based query language with logical operators and range filtering. The Python client of \textcite{rose2017} wraps this protocol but is scoped to AFLOW alone. OQMD exposes a RESTful HTTP interface through its \texttt{qmpy\_rester} wrapper \parencite{saal2013,kirklin2015}, which supports logical filter expressions and returns results as JSON. The lower-level \texttt{qmpy} library targets a locally replicated database rather than remote HTTP access.

datascribe\_api could not have been built by contributing to any of these existing libraries. Each database uses a different query language, returns a different response structure, and requires its own authentication flow, so a researcher drawing on all three must still install and maintain three separate packages and write format-normalization code before analysis can start. Extending \texttt{pymatgen} would cover Materials Project but not AFLOW, OQMD, or user-managed data tables; the AFLOW and OQMD clients cover only their own databases. datascribe\_api also needs to communicate with DataScribe's own REST API, which carries the data-table service, the FAIR-compliant metadata schema, and the provenance records DataScribe attaches at ingestion time. No existing database client provides any of these.

Experimental data is particularly hard to manage in collaborative or multi-group studies. Results arrive in formats that vary by group, instrument, and convention, and bringing them into a shared, queryable structure requires considerable effort before analysis can start. Within a single group, the absence of a structured repository means data is typically retrieved by re-reading files, which makes it hard to combine older measurements with new results or revisit filtering decisions. The difficulty compounds when the same data must feed downstream tasks such as surrogate model training or Bayesian optimization, since each task imposes its own format requirements. DataScribe provides a structured, FAIR-compliant store where groups can deposit and version experimental tables alongside curated database records. datascribe\_api exposes that store programmatically, so retrieved data flows directly into analysis and modeling workflows without project-specific parsing code.

\section{Software Design}

\begin{table}[t]
\centering
\footnotesize
\setlength{\tabcolsep}{7pt}
\renewcommand{\arraystretch}{1.35}
\caption{The nine endpoints exposed by datascribe\_api, grouped by functional domain. The seven data-table endpoints cover listing, row retrieval, filtered counting, and schema inspection for user-managed experimental tables. The two materials endpoints support cross-provider search and identifier-based lookup across the Materials Project, AFLOW, and OQMD. Required parameters are shown in bold and optional parameters in regular weight. Endpoints marked with $^\dagger$ accept \texttt{Filter} expressions.}
\label{tab:endpoints}
\begin{tabularx}{\linewidth}{ll>{\raggedright\arraybackslash}X}
\toprule
\textbf{Endpoint} & \textbf{Key Parameters} & \textbf{Description} \\
\midrule
\multicolumn{3}{l}{\textbf{Data tables}} \\[2pt]
\texttt{get\_data\_tables} & -- & List all tables (admin only) \\
\texttt{get\_data\_tables\_for\_user} & -- & List user-accessible tables \\
\texttt{get\_data\_table} & \textbf{tableName}, startingRow, numRows & Paginated row retrieval \\
\texttt{get\_data\_table\_rows}$^\dagger$ & \textbf{tableName}, \textbf{columns}, filters, numRows & Rows with column selection and filtering \\
\texttt{get\_data\_table\_rows\_count}$^\dagger$ & \textbf{tableName}, filters & Filtered row count \\
\texttt{get\_data\_table\_columns} & \textbf{tableName} & Column schema \\
\texttt{get\_data\_table\_metadata} & \textbf{tableName} & Table metadata \\[4pt]
\multicolumn{3}{l}{\textbf{Materials}} \\[2pt]
\texttt{get\_material\_by\_id} & \textbf{ids}, providers & Lookup by identifier \\
\texttt{search\_materials} & formula, elements, spacegroup, providers & Cross-provider search \\
\bottomrule
\end{tabularx}
\end{table}

Research scripts are typically short, written by domain scientists rather than software engineers, and run synchronously. An asynchronous client would offer higher throughput for concurrent requests but would require every calling script to adopt an asynchronous event loop, a pattern that is unfamiliar to many researchers and incompatible with standard Jupyter usage. datascribe\_api therefore uses the synchronous \texttt{requests} \parencite{reitz2011} library. The same reasoning favors explicit method arguments and typed return values over streaming or cursor-based APIs, which optimize for large-scale throughput at the cost of interface complexity.

A second trade-off concerns the filter API. Requiring users to construct filter dictionaries by hand is error-prone and verbose. The library instead lets researchers write filter conditions in standard Python comparison syntax; comparing a field name against a threshold produces the correct server-side predicate automatically. Filter expressions read naturally in notebooks and are easy to compose. Multiple conditions can be passed as a list; the library applies them with server-side AND logic.

A third trade-off is between a fixed set of named methods and an endpoint registry. The library derives its callable methods directly from a single route definition file, so new endpoints added to the DataScribe platform become immediately available in the client without any code changes. The resulting seven data-table endpoints and two materials-search endpoints (\autoref{tab:endpoints}) all appear in tab-completion in interactive shells and IDEs.

\begin{figure}[t]
\centering
\includegraphics[width=\linewidth]{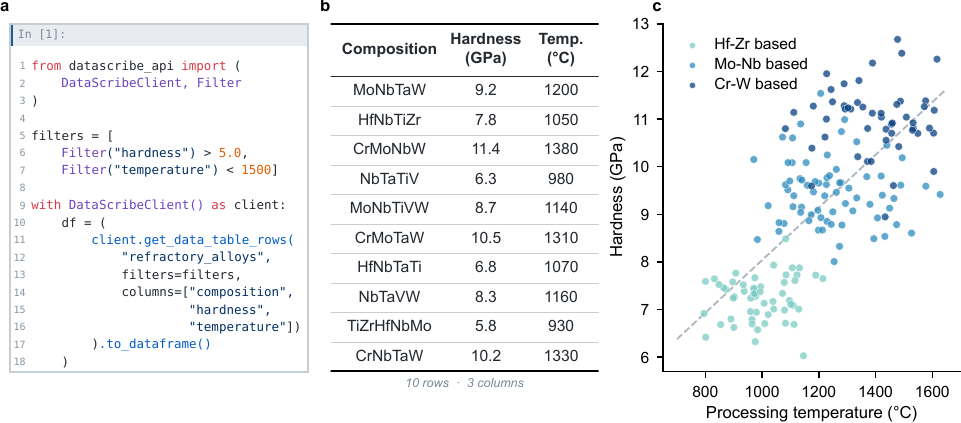}
\caption{End-to-end data retrieval workflow with datascribe\_api. In \textbf{a}, a Python code cell queries a custom data table using two filter conditions and column selection. Panel \textbf{b} shows the first ten rows of the pandas DataFrame returned by the query, and \textbf{c} presents a scatter plot of hardness versus processing temperature across three refractory high-entropy alloy families, generated from the retrieved data.}
\label{fig:workflow}
\end{figure}

All API responses are validated against typed Pydantic \parencite{pydantic2024} models at parse time, so schema mismatches surface as Python exceptions rather than silent incorrect values. Each result collection converts to a pandas DataFrame \parencite{mckinney2010,reback2020} in a single call, so retrieved data is directly available for statistical analysis or machine learning. The models are designed to remain compatible as the DataScribe API evolves: new fields added by the server are preserved rather than discarded, so existing client code continues to work without modification. Every materials search result includes a provenance field that identifies its source database, a requirement of FAIR data practice.

\subsection{Command-Line Interface}

The CLI mirrors all nine API endpoints as subcommands, using Typer \parencite{ramirez2019} for argument parsing and Rich \parencite{willmott2020} for terminal output. Every command supports a machine-readable JSON output mode for shell pipelines and LLM agents. Filter expressions are entered as plain strings on the command line and support the same twelve operators as the Python API.

The library is published on PyPI, archived on Zenodo with a persistent DOI \parencite{sariturk2025} for citation in publications, and pre-built for Python~3.10 and later on Linux, macOS, and Windows. Continuous integration runs the full test suite against every pull request. Documentation is at \url{https://datascribe-cloud.github.io/datascribe\_api/} and covers installation, all nine endpoints, CLI usage, and filter syntax with worked examples.

\section{Research Impact Statement}

DataScribe's closed-loop materials discovery workflows already use datascribe\_api as their data-access layer \parencite{singh2026}. In a documented case study, XGBoost regression on a refractory high-entropy alloy dataset retrieved through datascribe\_api achieved $R^2 = 0.956$ for creep merit prediction across values spanning multiple orders of magnitude. datascribe\_api also feeds DataScribe's Bayesian optimization and surrogate modeling pipelines through the same endpoints.

\autoref{fig:workflow} shows this end-to-end process. The filter conditions and column selection in \autoref{fig:workflow}\textbf{a} produce the pandas DataFrame shown in \autoref{fig:workflow}\textbf{b}, which feeds into the hardness-temperature scatter plot in \autoref{fig:workflow}\textbf{c}. Because the endpoint, filter conditions, column selection, pagination bounds, and provider list are all explicit Python values, a datascribe\_api script is itself a provenance record. This satisfies the traceability requirements of FAIR data practice without additional documentation.

The library's test suite has three modules. Each covers a distinct layer of the stack. The API module exercises all nine endpoints against the live service and verifies response types, required field presence, pagination behavior, and error handling for authentication failures, missing arguments, and requests for nonexistent resources. A filter module tests all twelve comparison operators and the serialization logic for single conditions, lists, and raw dictionary inputs. The CLI module tests each subcommand with valid and invalid inputs, JSON output mode, and the plain-string filter parser. Tests that query AFLOW are marked separately and excluded from the standard CI run, since AFLOW queries are rate-limited.

Reproducibility in materials machine learning depends as much on dataset construction as on model choice. Which alloy families are included, which property ranges are selected, which records pass a quality filter: these choices shape the training distribution, yet researchers rarely document them precisely when data is assembled through web interfaces or manual downloads. A datascribe\_api script makes dataset construction as explicit as model training code. The filtering logic that produced the training set is part of the published record, and any researcher with the same script and API access can reconstruct the dataset exactly.

\section{Acknowledgements}

R.A.\ and D.S.\ acknowledge support from the Army Research Office (ARO) through Grant No.\ W911NF-22-2-0117. Portions of this research were conducted with the advanced computing resources provided by Texas A\&M High Performance Research Computing.

\printbibliography

@article{singh2026,
  author    = {Singh, Divyanshu and Sar{\i}t{\"u}rk, Do{\u{g}}uhan and Lea, Cameron and Islam, Md Shafiqul and Arroyave, Raymundo and Attari, Vahid},
  title     = {{DataScribe: An AI-Native, Policy-Aligned Web Platform for Multi-Objective Materials Design and Discovery}},
  journal   = {arXiv preprint arXiv:2601.07966},
  year      = {2026},
  doi       = {10.48550/arXiv.2601.07966},
}

@article{Horton2025,
  author = {Horton,  Matthew K. and Huck,  Patrick and Yang,  Ruo Xi and Munro,  Jason M. and Dwaraknath,  Shyam and Ganose,  Alex M. and Kingsbury,  Ryan S. and Wen,  Mingjian and Shen,  Jimmy X. and Mathis,  Tyler S. and Kaplan,  Aaron D. and Berket,  Karlo and Riebesell,  Janosh and George,  Janine and Rosen,  Andrew S. and Spotte-Smith,  Evan W. C. and McDermott,  Matthew J. and Cohen,  Orion A. and Dunn,  Alex and Kuner,  Matthew C. and Rignanese,  Gian-Marco and Petretto,  Guido and Waroquiers,  David and Griffin,  Sinead M. and Neaton,  Jeffrey B. and Chrzan,  Daryl C. and Asta,  Mark and Hautier,  Geoffroy and Cholia,  Shreyas and Ceder,  Gerbrand and Ong,  Shyue Ping and Jain,  Anubhav and Persson,  Kristin A.},
  title = {{Accelerated data-driven materials science with the Materials Project}},
  journal = {Nature Materials},
  volume = {24},
  number = {10},
  pages = {1522–1532},
  year = {2025},
  doi = {10.1038/s41563-025-02272-0},
}

@article{jain2013,
  author    = {Jain, Anubhav and Ong, Shyue Ping and Hautier, Geoffroy and Chen, Wei and Richards, William Davidson and Dacek, Stephen and Cholia, Shreyas and Gunter, Dan and Skinner, David and Ceder, Gerbrand and Persson, Kristin A.},
  title     = {Commentary: {The Materials Project}: A materials genome approach to accelerating materials innovation},
  journal   = {APL Materials},
  volume    = {1},
  number    = {1},
  pages     = {011002},
  year      = {2013},
  doi       = {10.1063/1.4812323},
}

@article{curtarolo2012,
  author    = {Curtarolo, Stefano and Setyawan, Wahyu and Hart, Gus L. W. and Jahnatek, Michal and Chepulskii, Roman V. and Taylor, Richard H. and Wang, Shidong and Xue, Junkai and Yang, Kesong and Levy, Ohad and Mehl, Michael J. and Stokes, Harold T. and Demchenko, Denis O. and Morgan, Dane},
  title     = {{AFLOW}: An automatic framework for high-throughput materials discovery},
  journal   = {Computational Materials Science},
  volume    = {58},
  pages     = {218--226},
  year      = {2012},
  doi       = {10.1016/j.commatsci.2012.02.005},
}

@article{rose2017,
  author    = {Rose, Frisco and Toher, Cormac and Gossett, Eric and Oses, Corey and Nardelli, Marco Buongiorno and Fornari, Marco and Curtarolo, Stefano},
  title     = {{AFLOWLIB.org}: A distributed materials properties repository from high-throughput \textit{ab initio} calculations},
  journal   = {Computational Materials Science},
  volume    = {137},
  pages     = {362--370},
  year      = {2017},
  doi       = {10.1016/j.commatsci.2017.05.027},
}

@article{saal2013,
  author    = {Saal, James E. and Kirklin, Scott and Aykol, Muratahan and Meredig, Bryce and Wolverton, Christopher},
  title     = {{Materials Design and Discovery with High-Throughput Density Functional Theory: The Open Quantum Materials Database (OQMD)}},
  journal   = {JOM},
  volume    = {65},
  number    = {11},
  pages     = {1501--1509},
  year      = {2013},
  doi       = {10.1007/s11837-013-0755-4},
}

@article{kirklin2015,
  author    = {Kirklin, Scott and Saal, James E. and Meredig, Bryce and Thompson, Alex and Doak, Jeff W. and Aykol, Muratahan and R{\"u}hl, Stephan and Wolverton, Christopher},
  title     = {The Open Quantum Materials Database ({OQMD}): assessing the accuracy of {DFT} formation energies},
  journal   = {npj Computational Materials},
  volume    = {1},
  pages     = {15010},
  year      = {2015},
  doi       = {10.1038/npjcompumats.2015.10},
}

@article{taylor2014,
  author    = {Taylor, Richard H. and Rose, Frisco and Toher, Cormac and Levy, Ohad and Yang, Kesong and Nardelli, Marco Buongiorno and Curtarolo, Stefano},
  title     = {A {REST}ful {API} for exchanging materials data in the {AFLOWLIB.org} consortium},
  journal   = {Computational Materials Science},
  volume    = {93},
  pages     = {178--192},
  year      = {2014},
  doi       = {10.1016/j.commatsci.2014.05.014},
}

@article{ong2013,
  author    = {Ong, Shyue Ping and Richards, William Davidson and Jain, Anubhav and Hautier, Geoffroy and Kocher, Michael and Cholia, Shreyas and Gunter, Dan and Chevrier, Vincent L. and Persson, Kristin A. and Ceder, Gerbrand},
  title     = {Python Materials Genomics (pymatgen): A robust, open-source python library for materials analysis},
  journal   = {Computational Materials Science},
  volume    = {68},
  pages     = {314--319},
  year      = {2013},
  doi       = {10.1016/j.commatsci.2012.10.028},
}

@inproceedings{mckinney2010,
  author    = {McKinney, Wes},
  title     = {Data Structures for Statistical Computing in {Python}},
  booktitle = {{Proceedings of the 9th Python in Science Conference}},
  pages     = {56--61},
  year      = {2010},
  doi       = {10.25080/Majora-92bf1922-00a},
}

@software{reback2020,
  author    = {{The pandas development team}},
  title     = {pandas-dev/pandas: Pandas},
  year      = {2020},
  publisher = {Zenodo},
  doi       = {10.5281/zenodo.3509134},
  url       = {https://doi.org/10.5281/zenodo.3509134},
}

@software{pydantic2024,
  author    = {Colvin, Samuel and Jolibois, Eric and Ramsey, Hasan and Baig, Arshith and others},
  title     = {Pydantic},
  year      = {2024},
  publisher = {GitHub},
  url       = {https://github.com/pydantic/pydantic},
  version   = {2},
}

@software{reitz2011,
  author    = {Reitz, Kenneth},
  title     = {{Requests: HTTP for Humans}},
  year      = {2011},
  publisher = {GitHub},
  url       = {https://github.com/psf/requests},
}

@software{ramirez2019,
  author    = {Ram{\'i}rez, Sebasti{\'a}n},
  title     = {{Typer: Build great CLIs. Easy to code. Based on Python type hints}},
  year      = {2019},
  publisher = {GitHub},
  url       = {https://github.com/fastapi/typer},
}

@software{willmott2020,
  author    = {Willmott, Will},
  title     = {{Rich: Rich text and beautiful formatting in the terminal}},
  year      = {2020},
  publisher = {GitHub},
  url       = {https://github.com/Textualize/rich},
}

@software{sariturk2025,
  author    = {Sar{\i}t{\"u}rk, Do{\u{g}}uhan},
  title     = {datascribe\_api},
  year      = {2025},
  publisher = {Zenodo},
  doi       = {10.5281/zenodo.17090844},
  url       = {https://doi.org/10.5281/zenodo.17090844},
}

\end{document}